\documentclass[10pt,conference]{IEEEtran}
\IEEEoverridecommandlockouts
\usepackage{cite}
\usepackage{amsmath,amssymb,amsfonts}
\usepackage{algorithmic}
\usepackage{graphicx}
\usepackage{textcomp}
\usepackage{xcolor}
\def\BibTeX{{\rm B\kern-.05em{\sc i\kern-.025em b}\kern-.08em
    T\kern-.1667em\lower.7ex\hbox{E}\kern-.125emX}}

\usepackage[ruled,vlined]{algorithm2e}    
\usepackage{setspace}
\usepackage{algorithmic}
\usepackage{multirow}
\usepackage{comment}
\usepackage{tabularx}
\usepackage{booktabs}

\newcommand\norm[1]{\left\lVert#1\right\rVert}
\newcommand{\RomanNumeralCaps}[1]
    {\MakeUppercase{\romannumeral #1}}

\newcolumntype{L}{>{\centering\arraybackslash}m{4.5cm}}
\newcolumntype{K}{>{\centering\arraybackslash}m{2cm}}
\newcolumntype{R}{>{\centering\arraybackslash}m{4.5cm}}

\begin{document}

\title{Dynamic Range Improvement in\\ Bistatic Backscatter Communication\\ Using Distributed MIMO 
\thanks{©2023 IEEE. Personal use of this material is permitted. Permission from IEEE must be obtained for all other uses, in any current or future media, including reprinting/republishing this material for advertising or promotional	purposes, creating new collective works, for resale or redistribution to servers
or lists, or reuse of any copyrighted component of this work in other works.
	
DOI: 10.1109/GLOBECOM48099.2022.10000927
}}

\author{\IEEEauthorblockN{Ahmet Kaplan\IEEEauthorrefmark{1}, Joao Vieira\IEEEauthorrefmark{2}, and Erik G. Larsson\IEEEauthorrefmark{1}}
	\IEEEauthorblockA{\IEEEauthorrefmark{1}Department of Electrical Engineering (lSY), Linköping University, 581 83 Linköping, Sweden.\\
	\IEEEauthorrefmark{2}Ericsson Research, 223 62 Lund, Sweden.\\
	Email: \{ahmet.kaplan, erik.g.larsson\}@liu.se, joao.vieira@ericsson.com}}
\maketitle

\begin{abstract}
Backscatter communication (BSC) is a promising solution for Internet-of-Things (IoT) connections due to its low-complexity, low-cost, and energy-efficient solution for sensors. 
There are several network infrastructure setups that can be used for BSC with IoT nodes/passive devices. One of them is a bistatic setup where there is a need for high dynamic range and high-resolution analog-to-digital converters at the reader side.
In this paper, we investigate a bistatic BSC setup with multiple antennas. 
We propose a novel algorithm to suppress direct link interference between the carrier emitter (CE) and the reader using beamforming into the nullspace of the CE-reader direct link to decrease the dynamic range of the system and increase the detection performance of the backscatter device (BSD). 
Further, we derive a Neyman-Pearson (NP) test and an exact closed-form expression for its performance in the detection of the BSD. Finally, simulation results show that the dynamic range of the system is significantly decreased and the detection performance of the BSD is increased by the proposed algorithm compared to a system not using beamforming in the CE, which could then be used in a host of different practical fields such as agriculture, transportation, factories, hospitals, smart cities, and smart homes.
\end{abstract}


\section{Introduction}
There were almost 14.6 billion internet of things (IoT) connections in 2021. It is expected that the number of IoT connections will reach 30.2 billion by 2027 \cite{cerwall2021ericsson}. IoT devices have several usage areas such as in agriculture, transportation, factories, hospitals, smart cities, and smart homes. It is also desirable to have low-cost, low-complexity, and energy-efficient IoT devices. Moreover, with the emergence of 6G, the number of battery-less sensors/IoT devices will increase \cite{tariq2020speculative, chowdhury20206g}. Battery-less sensors can communicate by using backscatter communication (BSC). That is why BSC is a promising technique for massive connectivity in 6G \cite{chowdhury20206g}.

In a BSC setup, we usually have the following types of equipment: a carrier emitter (CE), a reader, and a backscatter device (BSD). The main BSC configurations are
\begin{itemize}
    \item \textbf{Monostatic:} In a monostatic BSC setup, the CE and reader are co-located and share many parts of the same infrastructure. For example, the CE and reader either share the same antenna elements or use separate antennas. The CE sends a radio frequency (RF) signal to the BSD, and the BSD modulates the incoming RF signal and backscatters it to the reader \cite{basharat2021reconfigurable}. The monostatic system suffers from round-trip path loss, and requires full-duplex technology if the same antennas are simultaneously used for transmission and reception.
    
    \item \textbf{Bistatic:} In a bistatic BSC setup, the CE and reader are spatially separated from each other, and therefore do not typically share RF circuitry, which is beneficial for a number of reasons \cite{kimionis2014increased}. For example, the CE and reader can be located separately to decrease the round-trip path loss. 
    
    \item \textbf{Ambient:} In an ambient BSC system, the CE and reader are separated in a similar way as in a bistatic system. However, the ambient system does not have a dedicated CE and uses ambient RF sources such as Bluetooth, Wi-Fi, or TV signals \cite{basharat2021reconfigurable}.
\end{itemize}

The system that exploits backscattering to transmit information was first proposed in \cite{stockman1948communication}. Recently, BSC has received significant attention from academia and industry with the emergence of massive IoT. In \cite{mishra2019optimal}, the authors optimize the number of orthogonal pilots, and the energy allocation for the channel estimation and information transmission phases in order to maximize the received signal-to-noise ratio at the reader in a monostatic system. They also show that the communication range of the monostatic BSC systems increases by using multiple antennas at the reader. Furthermore, it is shown in \cite{kashyap2016feasibility} that by adding more antennas to the reader in a monostatic system, we can extend the range for
wireless energy transfer while maintaining a given target outage probability. The authors of \cite{liu2014multi} investigate a monostatic system with multiple battery-less single-antenna users. They optimize time allocation and power allocation to maximize the minimum throughput for users under a perfect channel state information (PCSI) assumption.

In \cite{griffin2008gains}, the authors investigate the effect of the number of antennas in a BSD in a monostatic system under PCSI. They prove that we can increase the diversity gain by increasing the number of antennas in the BSD. As a result, the bit error rate (BER) performance is improved in the system. Furthermore, in \cite{boyer2013backscatter}, the diversity gain is investigated by changing the number of antennas at both the BSD and the reader under PCSI. They also show that the diversity order increases with the increasing number of antennas in the BSD.

In \cite{kimionis2014increased}, a bistatic system is investigated to achieve long-range BSC. The authors propose two modulation schemes: on-off keying (for the bandwidth-limited region) and frequency shift keying (for the power-limited region), and derive the receivers for both modulation schemes. They also present experimental results for the bistatic setup that includes a single antenna CE, reader, and BSD. They validate the long-range capability of bistatic links through experiments.

In a bistatic BSC, the backscattered signal is weak compared to the direct link interference due to the double path-loss effect on the two-way backscatter link. That is why the dynamic range of the reader, which is proportional to the signal strength ratio between the received signal from direct link and  the weak backscattered signal \cite{biswas2021direct}, can be high. As a result, a high resolution analog-to-digital converter (ADC) is required to detect the weak backscattered signal under heavy interference. In multiple antenna technology, high resolution ADCs are one of the major power consumers \cite{mollen2016uplink}. Moreover, the backscattered signal is pushed to the last bits of ADC due to the interference which causes a low signal-to-interference-plus-noise ratio (SINR) \cite{biswas2021direct}. 
Therefore, it would be beneficial if the direct link interference is suppressed before the automatic gain control (AGC) and ADC in order to decrease the dynamic range of the system.

The authors of \cite{biswas2021direct} investigate the coverage region for IoT communication, and the effect of the direct link interference on the dynamic range in the bistatic BSC and ambient BSC systems. They show that the high dynamic range limits the system performance.
In \cite{varshney2017lorea} and \cite{li2019capacity}, the carrier frequency of the reflected signal is changed at the BSD to solve the direct link interference problem in a single-input single-output (SISO) bistatic BSC system. In \cite{tao2021novel}, the authors apply Miller coding at the BSD and utilize the periodicity of the carrier signal to mitigate the direct link interference in a SISO bistatic BSC setup. However, the proposed method cancels the interference after the ADC which does not address the high-resolution ADC/high dynamic range problem.

\textbf{Our contribution:} In our paper, we investigate the bistatic setup with a multiantenna CE and reader. Our contributions can be summarized as follows:
\begin{itemize}
    \item To address the dynamic range problem in a bistatic BSC, we propose an algorithm to suppress the direct link interference by adapting the transmission from the CE using beamforming.
    \item We derive the optimal Neyman-Pearson (NP) test to detect the BSD/passive device.
    \item We provide an exact closed-form expression for the performance of the NP test, and analyze the performance of BSD detection at the reader in a bistatic BSC setup with multiple antennas.
\end{itemize}

The rest of this paper is organized as follows. In Section \ref{sec:system_model}, we introduce our system model. In Section \ref{sec:channel_estimation_ineterference_supression}, we give the details of our interference suppression algorithm. The NP detector is discussed in Section \ref{sec:detection_passive_device}. We present our simulation results in Section \ref{sec:numerical_results}. Finally, Section \ref{sec:conslision} concludes the paper.

\textbf{Notation:} In this paper, $(\cdot)^T$, $(\cdot)^*$, and $(\cdot)^H$ denote transpose, conjugate, and Hermitian transpose, respectively. $\operatorname{Re}\{\cdot\}$ and $\operatorname{Tr}\{\cdot\}$ denote the real part of a signal and trace of a matrix, respectively. We use $\norm{\cdot}$ for Frobenius norm. $E\{\cdot\}$ stands for the expected value. The italic, boldface capital, and boldface lowercase letters are used for scalars, matrices, and column vectors,  respectively.

\section{System Model} \label{sec:system_model}

\begin{figure}[tbp]
	\centering
	\includegraphics[width = 0.8\linewidth]{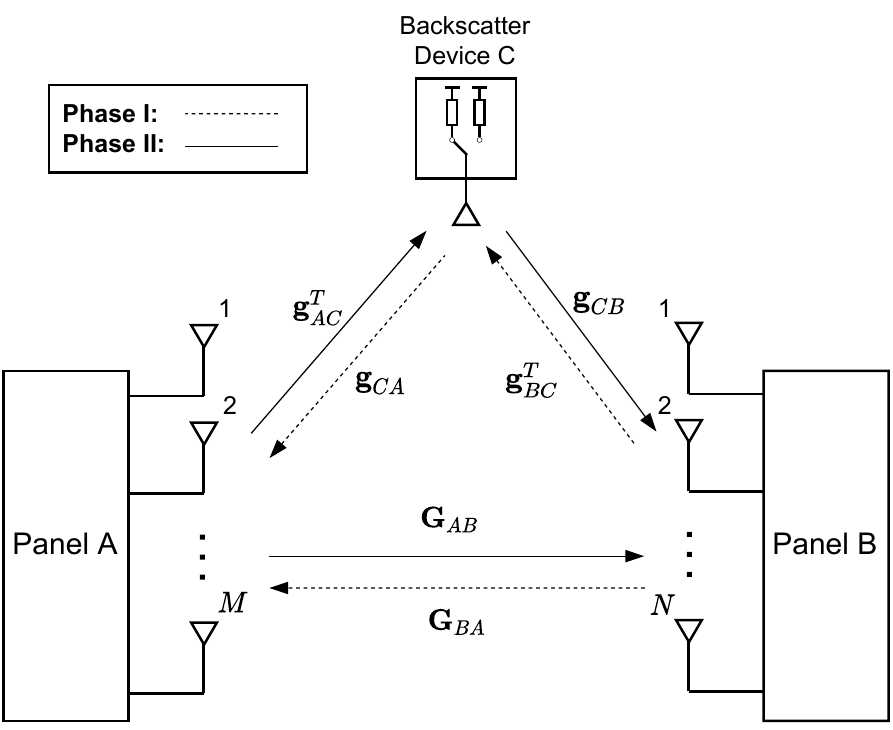}
	\caption{The system model of the multiantenna bistatic backscatter communication.}
	\label{fig:System_Model}
\end{figure}

In this section, we present the system model of our bistatic communication setup in Fig. \ref{fig:System_Model}. We also explain the channel estimation phase and BSD/passive device detection phase. 

In Fig. \ref{fig:System_Model}, Panel A (PanA) with $M$ antennas is the carrier emitter, Panel B (PanB) with $N$ antennas is the reader, and backscatter device C has a single antenna. The BSD can change its reflection coefficient by varying the impedance of the load connected to the BSD’s antenna to modulate the backscattered signal. Our aim is to decrease the dynamic range of the system, and detect the presence of the BSD. In Phase \RomanNumeralCaps{1} (P1), we estimate the channel between PanA and PanB. In Phase \RomanNumeralCaps{2} (P2), we construct a beamformer using a projection matrix which is designed based on the estimated channel to decrease the dynamic range and increase the detection performance by suppressing the interference due to the direct link $\text{PanA} \rightarrow \text{PanB}$.

In Fig. \ref{fig:System_Model}, $\mathbf{g}_{AC}^T, \mathbf{g}_{CA},  \mathbf{g}_{CB}, \mathbf{g}_{BC}^T, \mathbf{G}_{AB}$, and $\mathbf{G}_{BA}$ stand for the channels from PanA to BSD, BSD to PanA, BSD to PanB, PanB to BSD, PanA to PanB, and PanB to PanA, respectively. Due to the reciprocity, $\mathbf{G}_{AB} = \mathbf{G}_{BA}^T, \mathbf{g}_{AC}=\mathbf{g}_{CA}$ and $\mathbf{g}_{CB}=\mathbf{g}_{BC}$.
Here, the dimensions of $\textbf{g}_{AC}, \textbf{g}_{CB}$, and $\textbf{G}_{AB}$ are $M \times 1, N \times 1$, and $N \times M$, respectively.

\begin{figure*}[tb]
    \centering
    \includegraphics[width = 1\linewidth]{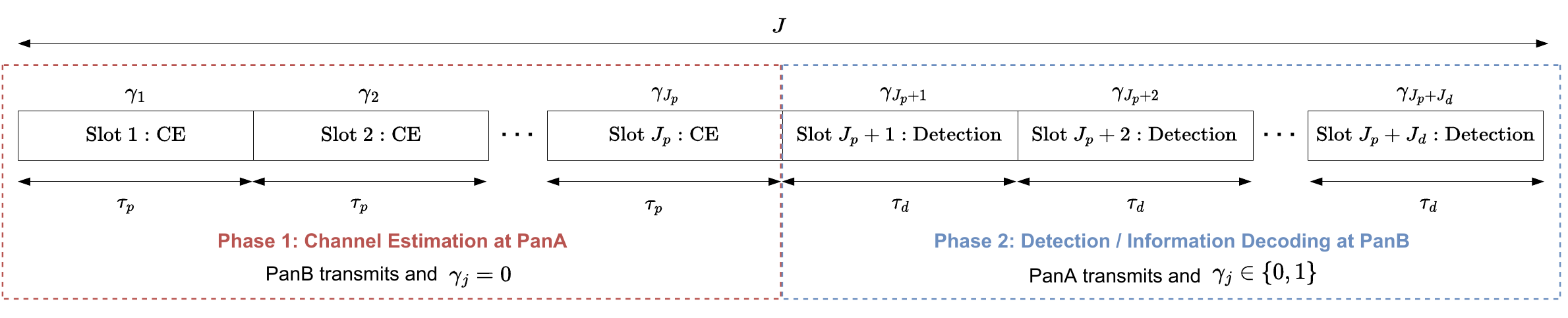}
    \caption{The transmission scheme of our bistatic setup.}
    \label{fig:TransmissionScheme}
\end{figure*}
In Fig. \ref{fig:TransmissionScheme}, the transmission scheme of our bistatic communication setup is given. There are two phases, as explained in the next subsections.

\subsection{Phase \RomanNumeralCaps{1}: Channel Estimation at PanA}
    The first phase comprises $J_p$ slots ($\tau_p J_p$ symbols). In each slot, PanB sends the same $N$ orthogonal pilot signals which each has $\tau_p$ symbols  in order to facilitate estimation of $\mathbf{G}_{BA}$ at PanA. The symbol length in seconds is denoted as $L$, so each orthogonal pilot signal has a length of $\tau_p L$ seconds. 
    The orthogonal pilot signals sent in a slot are written in matrix form as $\boldsymbol{\Phi} \in \mathbb{C}^{N \times \tau_p}$. Here, the pilot matrix $\boldsymbol{\Phi}$ satisfies $\boldsymbol{\Phi} \boldsymbol{\Phi}^{H} = \frac{p_t \tau_p L}{N}\mathbf{I}_N$ and $\tau_p \geq N$, where $p_t$ stands for the transmit power. The total amount of transmitted energy during P1 is expressed as $E_p  \triangleq J_p||\boldsymbol{\Phi}||^2 = J_p p_t \tau_p L$. 
    
    In P1, we select the reflection coefficients $\gamma_j=0$ for $j=1,2,\dotsc,J_p$, i.e., the BSD is silent in each slot. Here, $\gamma_j$ represents the relative reflection coefficient. When the BSD is silent, its reflection coefficient is a part of $\textbf{G}_{BA}$ like other scattering objects in the environment, and the contribution of the backscattered signal to the direct link is negligible. When $\gamma_j=1$, the relative difference in the channel is $\textbf{g}_{CA} \textbf{g}_{BC}^T$. It is also possible to design a BSD that absorbs the incoming signal for energy harvesting during $\gamma_j=0$ \cite{zawawi2018multiuser}.  Moreover, in the literature, several papers use on-off keying modulation with two states as used in this paper: $\gamma_j \neq 0$ and $\gamma_j=0$ \cite{hua2020bistatic, guo2018exploiting, zawawi2018multiuser}.

    \subsection{Phase \RomanNumeralCaps{2}: Detection at PanB}
    The second phase consists of $J_d$ slots ($\tau_d J_d$ symbols). In each slot, PanA sends a probing signal to detect the presence of the BSD at PanB. The probing signal sent in a slot is represented in matrix form as $\boldsymbol{\Psi} \in \mathbb{C}^{M \times \tau_d}$, where $\boldsymbol{\Psi} \boldsymbol{\Psi}^{H} = \alpha \textbf{I}_M = \frac{p_t \tau_d L}{M} \textbf{I}_M$, $\tau_d \geq M$. 
    
    The received signal of dimension $N \times \tau_d$ at PanB, in slot $j$ can be written as
\begin{equation} \label{eq:Phase2_InputOutputMatrixForm}
\mathbf{Y}_j=\mathbf{G}_{A B}\textbf{P}_s \boldsymbol{\Psi}+\gamma_j \mathbf{g}_{C B} \mathbf{g}_{A C}^{T}  \textbf{P}_s \boldsymbol{\Psi} + \mathbf{W}_j,
\end{equation}
where $j=J_p+1,J_p+2,\dotsc,J_p+J_d$ and $J_p+J_d=J$. 
We assume that, all the channels are time-invariant during the $J$ slot durations, i.e., the coherence time of all the channels exceeds $J$ slots. $\textbf{P}_s \in \mathbb{C}^{M \times M}$ is the scaled projection matrix. $\mathbf{W}_{j} \in \mathbb{C}^{N \times \tau_d}$ stands for the additive Gaussian noise and each element of $\textbf{W}_{j}$ is independent and identically distributed (i.i.d.) $\mathcal{CN}(0, 1)$.
$\gamma_j~\in~\{0,1\}$ denotes the reflection coefficients in the BSD, and $\gamma_j$ can be changed by varying the impedance of the load connected to the BSD's antenna. The dimensions of the quantities in the model are summarized in Table~1.
    \begin{table}[tbp]
	\caption{Parameters}
	\centering
	\label{tab:Parameters}
	\resizebox{0.42\textwidth}{!}{\begin{tabular}{|c|c|c|}
		\hline 
	\textbf{Parameter} & \textbf{Notation} & \textbf{Dimension} \\ \hline\hline
	    Received signal at PanB & $\textbf{Y}_j$  & $N \times \tau_d$ \\
	    \hline
	    Received signal at PanA & $\textbf{Y}_j^p$  & $M \times \tau_p$ \\ 
	    \hline
	    Channel from PanA to PanB & $\textbf{G}_{A B}$  & $N \times M$ \\ \hline
	    Channel from PanB to PanA & $\textbf{G}_{B A}$  & $M \times N$ \\ \hline
	    Channel from PanA to BSD & $\textbf{g}_{A C}^T$  & $1 \times M$ \\\hline
	    Channel from BSD to PanA & $\textbf{g}_{C A}$  & $M \times 1$ \\\hline
	    Channel from BSD to PanB & $\textbf{g}_{C B}$  & $N \times 1$ \\ \hline
	    Channel from PanB to PanB & $\textbf{g}_{B C}^T$  & $1 \times N$ \\ \hline
	    Projection matrix & $\textbf{P}$  & $M \times M$ \\ \hline
	    Probing signal & $\boldsymbol{\Psi}$  & $M \times \tau_d$ \\ \hline
	    Additive Gaussian noise at PanB & $\textbf{W}_j$  & $N \times \tau_d$ \\ 
	    \hline
	    Additive Gaussian noise at PanA & $\textbf{W}_j^p$  & $M \times \tau_p$ \\ \hline
	    Pilot signal & $\boldsymbol{\Phi}$  & $N \times \tau_p$ \\
	    \hline
	    Reflection coefficient at BSD & $\gamma_j$  & $1 \times 1$ \\ 
	    \hline
	\end{tabular}}
\vspace{-5pt}
\end{table}

In Eq. (\ref{eq:Phase2_InputOutputMatrixForm}), $\mathbf{G}_{A B}\textbf{P}_s \boldsymbol{\Psi}$ stands for the direct link interference.
The dynamic range of the system is
\begin{equation} \label{eq:DynamicRange}
	\zeta = E\left\{\frac{\left\|\mathbf{G}_{A B} \textbf{P}_s \boldsymbol{\Psi}\right\|^{2}+\left\|\mathbf{g}_{C B} \mathbf{g}_{A C}^{T} \textbf{P}_s \boldsymbol{\Psi}\right\|^{2}}{\left\|\mathbf{g}_{C B} \mathbf{g}_{A C}^{T} \textbf{P}_s \boldsymbol{\Psi}\right\|^{2}}\right\} ,
\end{equation}
where the expectation is taken with respect to random  channel estimation errors
(which affect $\textbf{P}_s$); the channels here are considered fixed.
When there is no projection, i.e., $\textbf{P}_s=\textbf{I}_M$, the dynamic range of the system can be large:  $\zeta \gg 1$.
When $\zeta \gg 1$, we need high-resolution ADCs which are not energy and cost efficient. Moreover, due to the high direct link interference, the system has low SINR, and the backscattered signal is pushed to the last bits of ADC. 
The scaled projection matrix, $\textbf{P}_s \in \mathbb{C}^{M \times M}$, to the nullspace of the dominant directions of $\textbf{G}_{AB}$ (or more exactly, an estimate of $\textbf{G}_{AB}$) is used to decrease the interference due to the direct link $\text{PanA} \rightarrow \text{PanB}$. As a result, the dynamic range of the system decreases. 

The next section explains our proposed algorithm to suppress the direct link interference and decrease the dynamic range.

\section{Proposed Interference Suppression Algorithm} \label{sec:channel_estimation_ineterference_supression}
In this section, we first present the channel estimation algorithm in P1 for the direct link between PanB and PanA. Then, we propose a novel algorithm based on the estimated direct link channel to mitigate the direct link interference at PanB in P2. The proposed algorithm decreases the dynamic range of the system, and increases the SINR and the detection performance. In practice, it also enables the use of low-resolution ADCs due to the decreased dynamic range.

\subsection{Channel Estimation at PanA}
In this subsection, we present the algorithm to estimate the channel from PanB to PanA, $\textbf{G}_{BA}$.

In the channel estimation phase, PanB sends the same pilot signal $\boldsymbol{\Phi}$ in each slot. At PanA, the received pilot signal in slot $j$ is given by
\begin{equation} \label{eq:Phase1OriginalEquation}
\textbf{Y}_j^p=\textbf{G}_{B A} \boldsymbol{\Phi} + \gamma_j \textbf{g}_{CA} \textbf{g}_{BC}^T \boldsymbol{\Phi} + \textbf{W}_j^p,
\end{equation}
where $j=1,2,\dots,J_p$ and $\textbf{W}_j^p \in \mathbb{C}^{M \times \tau_p}$ stands for the additive noise and all elements of $\textbf{W}_j^p$ are i.i.d. $\mathcal{CN}(0,1)$. We select the reflection coefficients
$\gamma_j=0$ for $j=1,2,\dotsc,J_p$, i.e., the BSD is silent. As a result, Eq. (\ref{eq:Phase1OriginalEquation}) can be simplified as 
\begin{equation}
    \textbf{Y}_j^p=\textbf{G}_{B A} \boldsymbol{\Phi} + \textbf{W}_j^p.
\end{equation}

The channel $\textbf{G}_{B A}$ is estimated by least-squares (LS) as follows:
\begin{equation}
    \hat{\textbf{G}}_{B A}=\frac{1}{J_p} \sum_{j=1}^{J_p} \mathbf{Y}_j^p \boldsymbol{\Phi}^H (\boldsymbol{\Phi} \boldsymbol{\Phi}^H)^{-1}.
\end{equation}
Due to the reciprocity,  the channel $\textbf{G}_{AB}$ is simply $\textbf{G}_{AB}=\textbf{G}_{BA}^T$;
the same holds for their estimates:  $\hat{\textbf{G}}_{A B}=\hat{\textbf{G}}_{B A}^T$. 

\subsection{Interference Suppression Algorithm} \label{section:interferenceSuppression}
This subsection explains our proposed algorithm to decrease the dynamic range and mitigate the direct link interference at PanB in P2. Based on the channel estimation in P1, PanA designs the projection matrix \textbf{P} to minimize the interference at PanB. The dynamic range of the system is decreased and the performance of the detection of BSD is improved at PanB. 

The singular value decomposition (SVD) of $\hat{\textbf{G}}_{A B}$ can be written as 
\begin{equation}\label{eq:svd}
    \hat{\textbf{G}}_{A B}=\textbf{U} \boldsymbol{\Delta} \textbf{V}^{\mathrm{H}},
\end{equation}
where $\textbf{U} \in \mathbb{C}^{N \times K_{0}}$ and $\textbf{V} \in \mathbb{C}^{M \times K_{0}}$ are semi-unitary matrices, and $K_{0} \leq \min\{M,N\}$ is the rank of $\hat{\textbf{G}}_{AB}$. $\boldsymbol{\Delta}$ is a $K_{0}\times K_{0}$ diagonal matrix with positive diagonal elements ordered in decreasing order.

In P2, PanA transmits a probing signal for PanB to detect the presence of the BSD. The probing signal, $\boldsymbol{\Psi}$, satisfies $\boldsymbol{\Psi} \boldsymbol{\Psi}^{H}=\alpha \textbf{I}_M$. Before transmitting $\boldsymbol{\Psi}$, PanA first designs a projection matrix, $\textbf{P}$, based on the channel estimation in P1 in order to decrease the interference at PanB due to the direct link $\text{PanA} \rightarrow \text{PanB}$. After that, PanA projects the probing signal onto the nullspace of $\hat{\textbf{G}}_{AB}$ in order to minimize the interference, decrease the dynamic range, and increase the detection probability of BSD at PanB. After the projection, PanA transmits the following signal
\begin{equation} \label{eq:projection}
	\textbf{P}_s \boldsymbol{\Psi} = \sqrt{\frac{M}{M-K}} \textbf{P} \boldsymbol{\Psi},
\end{equation}
where 
$\textbf{P} $ is an orthogonal projection of dimension $M \times M$ and rank $M-K$, with $K$ to be appropriately selected. 
We select $\textbf{P}$ to project onto the orthogonal complement of the  space spanned by the columns of $\textbf{V}_{K}$:
\begin{equation}
	\textbf{P}=\textbf{I}-\textbf{V}_{K} \textbf{V}_{K}^{\mathrm{H}},
\end{equation}
where $\textbf{V}_{K}$ contains the first $K$ columns of $\textbf{V}$ in Eq. (\ref{eq:svd}).
A suitable choice of $K$ is the dimension of the row space of $\textbf{G}_{AB}$ in order to project the probing signal (mostly) onto the nullspace of $\textbf{G}_{AB}$.
In Eq. (\ref{eq:projection}), $\sqrt{\frac{M}{M-K}}$ is used to keep the total radiated energy the same during P2 after the projection as follows $E_d  \triangleq J_d||\boldsymbol{\Psi}||^2 = J_d||\textbf{P}_s\boldsymbol{\Psi}||^2$.
Note that, the proposed algorithm works for a triangle setup, i.e., $\textbf{g}_{C B} \textbf{g}_{A C}^{T} \neq k \textbf{G}_{A B}$, for all $k \in \mathbb{C}$.

In summary, the received signals during both phases are given by 
\begin{subequations} 
    \begin{align}
	&\textbf{Y}_j^p=\textbf{G}_{B A} \boldsymbol{\Phi} + \textbf{W}_j^p, j \in \{1,\dotsc,J_p\} \\
		&\textbf{Y}_j=\textbf{G}_{A B}\textbf{P}_s \boldsymbol{\Psi}+\gamma_j \textbf{g}_{C B} \textbf{g}_{A C}^{T}  \textbf{P}_s \boldsymbol{\Psi} + \textbf{W}_j, j \in \{J_p+1,\dotsc,J\}
\label{eq:jam_coef_uncoded_1}
\end{align}
\end{subequations} 
\vspace{-15px}
\section{Detection of the Passive Device} \label{sec:detection_passive_device}
\vspace{-2px}
In this section, we provide the optimal NP test and its performance for the detection of the BSD in P2. The detection of the BSD at PanB is also equivalent to a binary modulation scheme with the alphabet which has two values $ `` 0"$ and $ `` 1"$. This problem is formulated as a hypothesis testing problem where the scenarios conditioned on the absence and presence of the BSD are taken as $\mathcal{H}_{0}$ and $\mathcal{H}_{1}$, respectively:
\begin{equation} \label{eq:hypothesisTesting}
\begin{split}
\mathcal\mathcal\mathcal{H}_{0}&:  \textbf{Y}_j=\textbf{G}_{A B} \textbf{P}_s \boldsymbol{\Psi} + \textbf{W}_j\\
\mathcal\mathcal\mathcal{H}_{1}&: \textbf{Y}_j =\textbf{G}_{A B} \textbf{P}_s \boldsymbol{\Psi} +\gamma_j \textbf{g}_{C B} \textbf{g}_{A C}^{T}  \textbf{P}_s \boldsymbol{\Psi}+ \textbf{W}_j,
\end{split}
\end{equation}
where $j \in \mathcal{S} = \{J_{p}+1,\dotsc, J\}$. 
We assume that the receiver have PCSIs, i.e., $\textbf{G}_{A B}$, $\textbf{g}_{C B}$ and $\textbf{g}_{A C}$ are known. We also assume that $\boldsymbol{\Psi}$ and $\textbf{P}_s$ are known by the receiver.

We apply the optimal NP test to detect the presence of the BSD as follows:
\vspace{-5px}
\begin{equation}\label{eq:NP}
    L = \frac{ 
    \prod_{j\in\mathcal{S}} p\left(\mathbf{Y}_j \mid \mathcal{H}_{1}\right)}
    {\prod_{j\in\mathcal{S}} p\left(\mathbf{Y}_j \mid \mathcal{H}_{0}\right)} \underset{\mathcal{H}_{0}}{\overset{\mathcal{H}_{1}}{\gtrless}} \eta,
\end{equation}
where $\eta$ is the detection threshold. Here, the NP detector will decide $\mathcal{H}_{1}$, if the likelihood ratio, $L$, is bigger than $\eta$. $p\left(\mathbf{Y}_j \mid \mathcal{H}_{1}\right)$ and $p\left(\mathbf{Y}_j \mid \mathcal{H}_{0}\right)$ denote the probability density functions (pdf) of the observations under $\mathcal{H}_{1}$ and $\mathcal{H}_{0}$, respectively, and are given as
\begin{subequations} 
\allowdisplaybreaks
    \begin{align}
    &p\left(\mathbf{Y}_j \mid \mathcal{H}_{1}\right) = \nonumber\\
    &\quad \frac{1}{\pi^{N \tau_d}} \exp \left[-||\textbf{Y}_j-\textbf{G}_{A B} \textbf{P}_s \boldsymbol{\Psi} - \gamma_j \textbf{g}_{C B} \textbf{g}_{A C}^{T}  \textbf{P}_s \boldsymbol{\Psi}||^2\right], \\
    &p\left(\mathbf{Y}_j \mid \mathcal{H}_{0}\right) = \frac{1}{\pi^{N \tau_d}} \exp \left[-||\textbf{Y}_j-\textbf{G}_{A B} \textbf{P}_s \boldsymbol{\Psi}||^2\right].
\end{align}
\end{subequations}
Let define $\textbf{A}_j= \gamma_j \textbf{g}_{CB} \textbf{g}_{AC}^T \textbf{P}_s \boldsymbol{\Psi}$ and $\textbf{Y}_j^\prime= \textbf{Y}_j - \textbf{G}_{AB} \textbf{P}_s \boldsymbol{\Psi}$. We can write $\log(L)$ as follows:
\begin{subequations}
\allowdisplaybreaks
\begin{align}
   \log(L) &= -\sum_{j \in \mathcal{S}} ||\textbf{Y}_j-\textbf{G}_{A B} \textbf{P}_s \boldsymbol{\Psi} - \gamma_j \textbf{g}_{C B} \textbf{g}_{A C}^{T}  \textbf{P}_s \boldsymbol{\Psi}||^2
   \nonumber \\
   &\qquad +\sum_{j \in \mathcal{S}} ||\textbf{Y}_j-\textbf{G}_{A B} \textbf{P}_s \boldsymbol{\Psi}||^2 \\
   &= -\sum_{j \in \mathcal{S}} ||\textbf{Y}_j^\prime - \textbf{A}_j||^2 + \sum_{j \in \mathcal{S}} ||\textbf{Y}_j^\prime||^2 \\
   &= \sum_{j \in \mathcal{S}}  \left(2\operatorname{Re}\{\operatorname{Tr}\{\textbf{A}_j {\textbf{Y}_j^\prime}^H\}\} - ||\textbf{A}_j||^2\right)
\end{align}
\end{subequations}
Finally, we can write the optimal NP test as follows:
\begin{equation}
\label{eq:FinalGLRT}
L^\prime = \sum_{j \in \mathcal{S}}  \operatorname{Re}\{\operatorname{Tr}\{\textbf{A}_j {\textbf{Y}_j^\prime}^H\}\}  \underset{\mathcal{H}_{0}}{\overset{\mathcal{H}_{1}}{\gtrless}} \eta^\prime,
\end{equation}
where $L^\prime=\frac{\log(L)+\sum_{j \in \mathcal{S}}||\textbf{A}_j||^2}{2}$ and $\eta^\prime=\frac{\log(\eta)+\sum_{j \in \mathcal{S}}||\textbf{A}_j||^2}{2}$.

To determine the detection performance of the NP test, we first find the distribution of the test statistic, $L^\prime$, both under $\mathcal{H}_{1}$ and $\mathcal{H}_{0}$. We consider that $\textbf{G}_{A B}, \textbf{g}_{C B},  \textbf{g}_{A C}$ and $\textbf{P}_s$ are deterministic.
Under $\mathcal{H}_{1}$, the test statistic is given by
\begin{subequations}
\allowdisplaybreaks
\begin{align}
    L^\prime&=\sum_{j \in \mathcal{S}} \operatorname{Re}\{\operatorname{Tr}\{\textbf{A}_j (\textbf{A}_j+\textbf{W}_j)^H\}\} \\
    &=\sum_{j \in \mathcal{S}} \big(\operatorname{Re}\{\operatorname{Tr}\{\textbf{A}_j \textbf{W}_j^H\}\} + ||\textbf{A}_j||^2\big)\\
    &=\sum_{j \in \mathcal{S}} \left(\operatorname{Re}\left\{\sum_{k=1}^N \sum_{l=1}^{\tau_d}\textbf{A}_{jkl} \textbf{W}_{jkl}^*\right\} + ||\textbf{A}_j||^2\right),
\end{align}
\end{subequations}
where $\textbf{A}_{jkl}$ and $\textbf{W}_{jkl}^*$ denote the element in the $k$-th row and $l$-th column in $\textbf{A}_{j}$ and $\textbf{W}_{j}^*$, respectively. $\textbf{A}_{jkl} \textbf{W}_{jkl}^*$ is i.i.d. $\mathcal{CN}(0, |\textbf{A}_{jkl}|^2)$, and the distribution of the test statistic under $\mathcal{H}_{1}$ is
\begin{equation}
L^\prime \sim \mathcal{N}(\sum_{j \in \mathcal{S}}||\textbf{A}_{j}||^2, \frac{1}{2}\sum_{j \in \mathcal{S}}||\textbf{A}_{j}||^2).
\end{equation}
Under $\mathcal{H}_{0}$, the test statistic is given by
\begin{equation}
    L^\prime = \sum_{j \in \mathcal{S}} \operatorname{Re}\{\operatorname{Tr}\{\textbf{A}_j \textbf{W}_j^H\}\} \sim \mathcal{N}(0, \frac{1}{2}\sum_{j \in \mathcal{S}}||\textbf{A}_{j}||^2).
\end{equation}
In summary,
\begin{equation}
    L^\prime \sim \begin{cases}\mathcal{N}(\sum_{j \in \mathcal{S}}||\textbf{A}_{j}||^2, \frac{1}{2}\sum_{j \in \mathcal{S}}||\textbf{A}_{j}||^2) & \text { under } \mathcal{H}_{1} \\ \mathcal{N}(0, \frac{1}{2}\sum_{j \in \mathcal{S}}||\textbf{A}_{j}||^2) & \text { under } \mathcal{H}_{0} .\end{cases}
\end{equation}

The probability of detection $(P_D)$ and the probability of false alarm $(P_{FA})$ are calculated as 
\begin{subequations} \label{eq:Pfa_Pd}
\allowdisplaybreaks
\begin{align}
    P_D &= \operatorname{Pr}\{L^\prime>\eta^\prime; \mathcal{H}_{1}\} 
    = Q\left( \frac{\eta^\prime-\sum_{j \in \mathcal{S}}||\textbf{A}_{j}||^2}{\sqrt{\frac{1}{2}\sum_{j \in \mathcal{S}}||\textbf{A}_{j}||^2}} \right), \\
    P_{FA} &= \operatorname{Pr}\{L^\prime>\eta^\prime; \mathcal{H}_{0}\}  
    = Q\left( \frac{\eta^\prime}{\sqrt{\frac{1}{2}\sum_{j \in \mathcal{S}}||\textbf{A}_{j}||^2}} \right). \label{eq:P_FA}
\end{align}
\end{subequations}

Finally, we can write the following relation between $P_D$ and $P_{FA}$ as follows:
\begin{equation}\label{eq:theoreticalNP}
    P_D = Q\left(Q^{-1}(P_{FA})-\sqrt{2\sum_{j \in \mathcal{S}}||\textbf{A}_{j}||^2}\right).
\end{equation}

\section{Numerical Results} \label{sec:numerical_results}
In this section, we first provide the simulation parameters and then discuss numerical results. 
$\textbf{g}_{AC}^T, \textbf{g}_{CA},  \textbf{g}_{CB}, \textbf{g}_{BC}^T, \textbf{G}_{AB}$, and $\textbf{G}_{BA}$ are modeled as line-of-sight channels with the path-loss coefficients
\begin{subequations}
	\allowdisplaybreaks
	\begin{align}
		\beta_{A C}&= \beta_{C A} = \frac{1}{d_{A C}^2}, \\
		\beta_{C B}&= \beta_{B C} = \frac{1}{d_{C B}^2}, \\
		\beta_{A B}&= \beta_{B A} = \frac{1}{d_{A B}^2},
	\end{align}
\end{subequations}
respectively \cite[Sec. 7.2]{tse2005fundamentals}. 
Here, $d_{A C},d_{C B},$ and $d_{A B}$ stand for the distances in meters between PanA-BSD, BSD-PanB, and PanA-PanB, respectively. 
We choose a uniform linear array at both PanA and PanB. 

\begin{table}[tbp]
	\caption{Simulation Parameters}
	\centering
	\label{tab:simulationParameters}
	\resizebox{0.480\textwidth}{!}{\begin{tabular}{|L|R|}
			\hline 
			\textbf{Parameter} &  \textbf{Value} \\ \hline\hline
			Symbol length in seconds & $L=5\times10^{-6}$ \\ \hline
			Number of slots for the probing signal & $J_d=2$ \\ \hline
			Number of symbols in each slot for the probing signal & $\tau_d=16$ \\ \hline
			Number of slots for the pilot signal & $J_p=1$ \\ \hline
			Number of symbols in each slot for the pilot signal  & $\tau_p=16$ \\ \hline		
			Number of antennas in PanA & $M=16$ \\ \hline
			Number of antennas in PanB & $N=16$ \\ \hline
			Number of antennas in BSD & $1$ \\ \hline
			The location of PanA, PanB, and BSD in meters & $(0,0), (6,0)$, and $(3,y),$ where $y \in [0\ 20]$  \\ \hline
			Inter-antenna distance both in PanA and PanB in meter & $0.5 \lambda=0.05$ \\ \hline
			Noise variance & $1$ \\ \hline
			Reflection coefficients at BSD & $\gamma_j = 0$ for $j=1,\dotsc,J_p$ and $\gamma_j \in \{0,1\}$ for $j \in \mathcal{S}$ \\ \hline
			First $K$ columns of \textbf{V} & $K=1$ \\ \hline
			SNR during the detection of BSD & $\text{SNR}_d=-10 \text{ dB}$ \\
			\hline
			SNR during the channel estimation & $\text{SNR}_p=0-30 \text{ dB}$ \\
			\hline
	\end{tabular}}
\end{table}

We use the following parameters in all simulations: $L=5\times10^{-6}$ \cite[Ch. 8]{dobkin2012rf}, $J_d=2, \tau_d=16, J_p=1, \tau_p=16, M=16, N=16, K=1$ and $\lambda=0.1 \text{ m},$ where $\lambda$ denotes the wavelength of the emitted signal. We select $K=1$ to design our projection matrix, because $\textbf{G}_{AB}$ is a rank-1 matrix.
We select the reflection coefficients at the BSD as $\gamma_j = 0$ for $j=1,\dotsc,J_p$ in P1 and $\gamma_j \in \{0,1\}$ for $j \in \mathcal{S}$ in P2, and we have the same number of $\gamma_j=0$ and $\gamma_j=1$ in P2.
The signal-to-noise ratio (SNR) during the channel estimation part is defined as $\text{SNR}_p=\beta_{BA} p_t J_p \tau_p L$, where $p_t \text{ and } L$ are the transmit power and the symbol length in seconds. The SNR during the detection of BSD is given as $\text{SNR}_d=\beta_{CB} \beta_{AC} p_t J_d \tau_d L \bar{\gamma}$, where $\bar{\gamma}=0.5$ is the average value of the reflection coefficients in P2.
PanA and PanB are located at $(0,0)$ and $(6,0)$ in meters, respectively. We list all the simulation parameters in Table \ref{tab:simulationParameters}.

\begin{figure}[tbp]
	\centering
	\includegraphics[width = 0.95\linewidth]{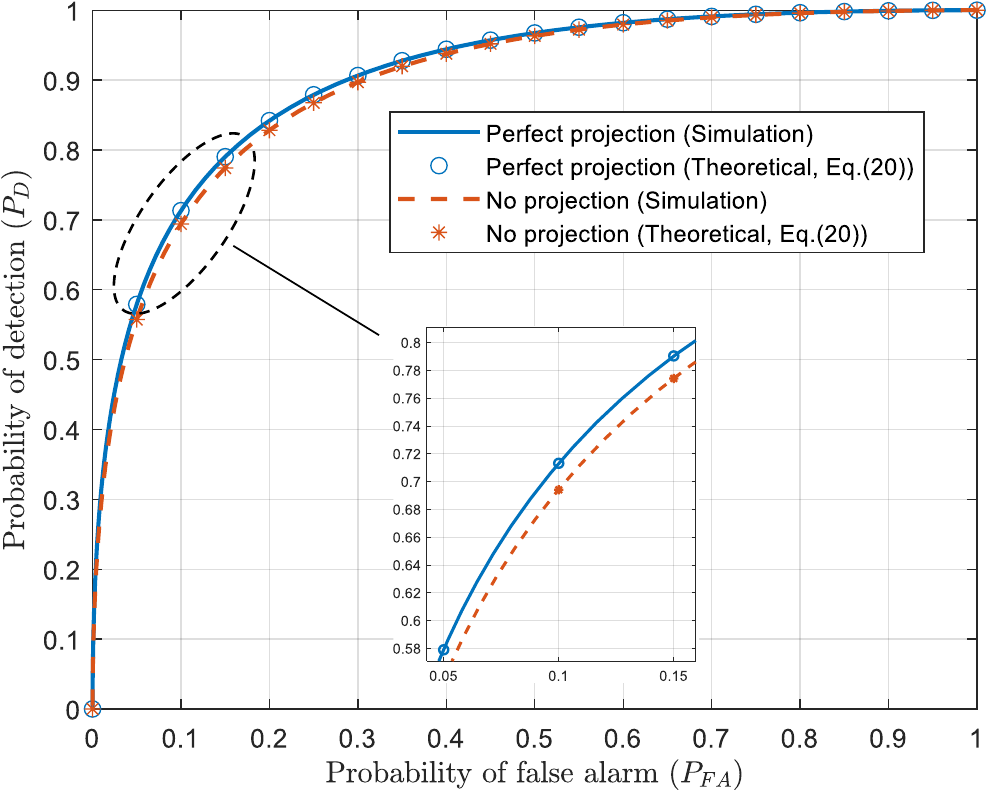}
	\caption{The detection performance of BSD at PanB.}
	\label{fig:ROC}
\end{figure}

In Fig. \ref{fig:ROC}, the theoretical and simulation results for the detection performance of the optimal NP test in Eq. (\ref{eq:FinalGLRT}) are given.
We used $10^5$ Monte-Carlo runs for the simulation results.
A triangular setup is used and the BSD is located at $(3,3)$ in meters. We select $\text{SNR}_d=-10 \text{ dB}$, and consider two different scenarios: (1) perfect projection and (2) no projection to investigate the effect of the scaled projection matrix $\textbf{P}_s$ on $P_D$ and $P_{FA}$. While $\textbf{P}_s=\textbf{I}_M$ for the no-projection case, the projection matrix is designed based on PCSI, i.e., $\hat{\textbf{G}}_{AB}=\textbf{G}_{AB}$, by PanA for the perfect-projection case. Since the projection matrices in both scenarios are deterministic, we use Eq. (\ref{eq:theoreticalNP}) to calculate the theoretical curves. In the figure, the theoretical results exactly match with the computer simulation results.
\begin{figure}[tbp]
    \centering
    \includegraphics[width = .95\linewidth]{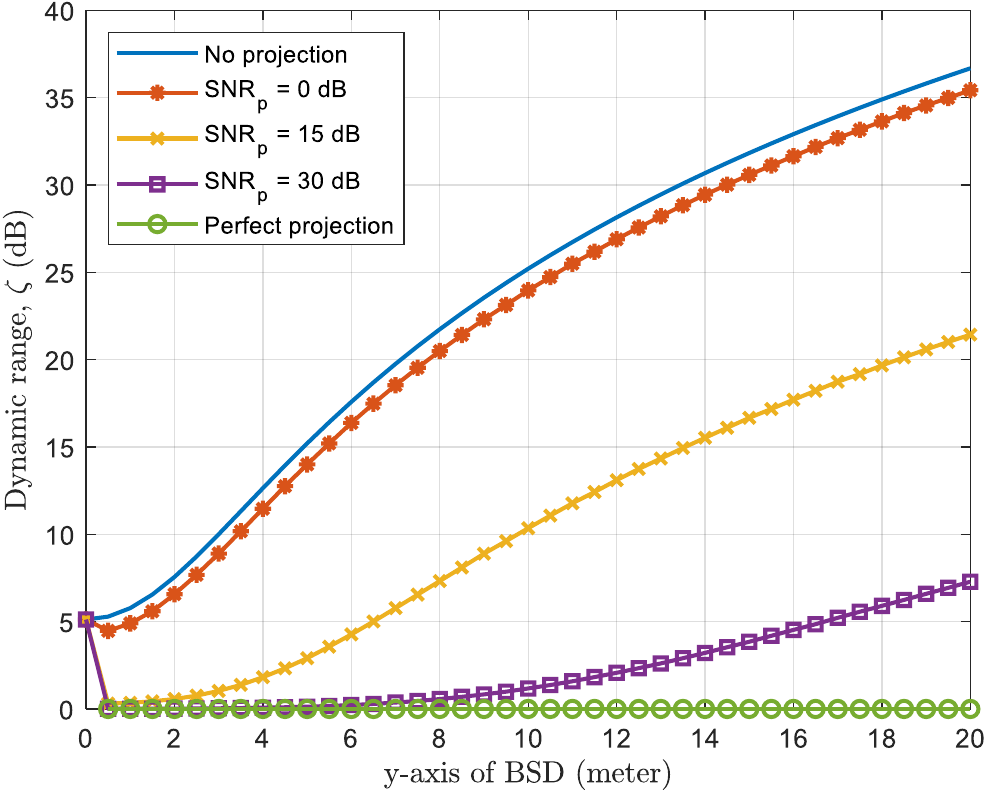}
    \caption{The ratio of the backscatter plus direct link power to the backscatter link power.}
    \label{fig:DynamicRange}
\end{figure}

The system which works with perfect projection matrix projects the transmitted signal to the nullspace of the dominant direction of $\textbf{G}_{AB}$. As seen in the figure, the detection performance is better in the case of perfect projection. It is observed because the radiated power in the directions which are close to the dominant direction of $\textbf{G}_{AB}$ decreases and the emitted power in all other directions increases. As a result, the received power at PanB from BSD signal increases with an accurate projection matrix. For example, at $P_{FA}=0.1$ in Fig. \ref{fig:ROC}, the system with perfect projection has almost $0.02$ gain in $P_{D}$ when compared to the system with no projection.

In Fig. \ref{fig:DynamicRange}, we sketch the ratio $\zeta$ given in Eq. (\ref{eq:DynamicRange}) for different BSD locations to investigate the change in the dynamic range of the system with and without projection. 
PanA and PanB are located at $(0,0)$ and $(6,0)$ in meters, respectively and BSD is located at $(3,y)$, where we change the vertical position of BSD, $y$, between $0$ and $20$ meters. Here, PanA estimate the direct channel using the transmitted pilot signal by PanB in P1. 
We use $10^3$ different realizations of the estimated direct channel to obtain Monte-Carlo simulation results.
We compare $\zeta$ for varying  $\text{SNR}_p$ values from $0$ dB to $30$ dB.  As seen in the figure, at $y=0$, the projection matrix does not affect $\zeta$, because there exists a complex number $k$ such that $\textbf{g}_{C B} \textbf{g}_{A C}^{T} = k \textbf{G}_{A B}$. In addition, at $y=0$, $\zeta$ is not defined in the case of perfect projection as seen in Eq. (\ref{eq:DynamicRange}). 

In Fig. \ref{fig:DynamicRange}, the dynamic range, $\zeta$, decreases with increasing $\text{SNR}_p$ when $y>0$ m. This is because the projection matrix designed with high $\text{SNR}_p$ values has more capability to decrease the interference due to the direct link $\text{PanA} \rightarrow \text{PanB}$. 
For example, at $y=10 \text{ m}$, $\zeta$ is $25.19 \text{ dB}, 23.95\text{ dB}, 10.34\text{ dB}, 1.17\text{ dB},$ and $0\text{ dB}$ for the no projection, $\text{SNR}_p=0$ dB, $\text{SNR}_p=15$ dB, $\text{SNR}_p=30$ dB, and perfect projection cases, respectively. 
In practice, with a decreased dynamic range, it is possible to use low-resolution ADCs which are more cost and energy efficient than high-resolution ADCs.

\section{Conclusion} \label{sec:conslision}
We present a novel algorithm to suppress the direct link interference at the reader in a bistatic BSC setup with multiple antennas. Our algorithm first estimates the channel between the CE (PanA) and the reader (PanB). Using the estimated channel, we design a projection matrix which is then used as a beamformer in order to mitigate the direct link interference and decrease the required dynamic range of the reader. We show that the dynamic range of the system is significantly decreased by using the projection matrix. That allows us to use low-resolution ADCs which are low-cost and more energy-efficient than high-resolution ones. Further, we derive the NP test and a closed-form expression for its performance for
the detection of the BSD at the reader. We show that the detection performance of BSD at the reader increases with an accurate projection matrix in a triangle setup. Moreover, a detector design without PCSI at the reader is left for future work.

\section*{Acknowledgement}
This work was funded by the REINDEER project of the European Union's Horizon 2020 research and innovation program under grant agreement No. 101013425, and in part by ELLIIT and the KAW foundation.

\bibliographystyle{IEEEtran}
\bibliography{references}

\begin{thebibliography}{10}
\providecommand{\url}[1]{#1}
\csname url@samestyle\endcsname
\providecommand{\newblock}{\relax}
\providecommand{\bibinfo}[2]{#2}
\providecommand{\BIBentrySTDinterwordspacing}{\spaceskip=0pt\relax}
\providecommand{\BIBentryALTinterwordstretchfactor}{4}
\providecommand{\BIBentryALTinterwordspacing}{\spaceskip=\fontdimen2\font plus
\BIBentryALTinterwordstretchfactor\fontdimen3\font minus
  \fontdimen4\font\relax}
\providecommand{\BIBforeignlanguage}[2]{{%
\expandafter\ifx\csname l@#1\endcsname\relax
\typeout{** WARNING: IEEEtran.bst: No hyphenation pattern has been}%
\typeout{** loaded for the language `#1'. Using the pattern for}%
\typeout{** the default language instead.}%
\else
\language=\csname l@#1\endcsname
\fi
#2}}
\providecommand{\BIBdecl}{\relax}
\BIBdecl

\bibitem{cerwall2021ericsson}
P.~Cerwall \emph{et~al.}, ``{The Ericsson mobility report},'' Ericsson,
  Stockholm, Sweden, White Paper EAB-21:010887 Uen, Dec. 2021.

\bibitem{tariq2020speculative}
F.~Tariq, M.~R. Khandaker, K.-K. Wong, M.~A. Imran, M.~Bennis, and M.~Debbah,
  ``{A speculative study on 6G},'' \emph{IEEE Wireless Commun.}, vol.~27,
  no.~4, pp. 118--125, Aug. 2020.

\bibitem{chowdhury20206g}
M.~Z. Chowdhury, M.~Shahjalal, S.~Ahmed, and Y.~M. Jang, ``{6G wireless
  communication systems: Applications, requirements, technologies, challenges,
  and research directions},'' \emph{IEEE Open J. Commun. Soc.}, vol.~1, pp.
  957--975, Jul., 2020.

\bibitem{basharat2021reconfigurable}
S.~Basharat, S.~A. Hassan, A.~Mahmood, Z.~Ding, and M.~Gidlund,
  ``{Reconfigurable intelligent surface-assisted backscatter communication: A
  new frontier for enabling 6G IoT networks},'' \emph{arXiv preprint
  arXiv:2107.07813}, 2021.

\bibitem{kimionis2014increased}
J.~Kimionis, A.~Bletsas, and J.~N. Sahalos, ``Increased range bistatic scatter
  radio,'' \emph{IEEE Trans. Commun.}, vol.~62, no.~3, pp. 1091--1104, Mar.
  2014.

\bibitem{stockman1948communication}
H.~Stockman, ``Communication by means of reflected power,'' \emph{Proc. IRE},
  vol.~36, no.~10, pp. 1196--1204, Oct. 1948.

\bibitem{mishra2019optimal}
D.~Mishra and E.~G. Larsson, ``{Optimal channel estimation for
  reciprocity-based backscattering with a full-duplex MIMO reader},''
  \emph{IEEE Trans. Signal Process.}, vol.~67, no.~6, pp. 1662--1677, Mar.
  2019.

\bibitem{kashyap2016feasibility}
S.~Kashyap, E.~Bj{\"o}rnson, and E.~G. Larsson, ``On the feasibility of
  wireless energy transfer using massive antenna arrays,'' \emph{IEEE Trans.
  Wireless Commun.}, vol.~15, no.~5, pp. 3466--3480, May 2016.

\bibitem{liu2014multi}
L.~Liu, R.~Zhang, and K.-C. Chua, ``Multi-antenna wireless powered
  communication with energy beamforming,'' \emph{IEEE Trans. Commun.}, vol.~62,
  no.~12, pp. 4349--4361, Dec. 2014.

\bibitem{griffin2008gains}
J.~D. Griffin and G.~D. Durgin, ``{Gains for RF tags using multiple
  antennas},'' \emph{IEEE Trans. Antennas Propag.}, vol.~56, no.~2, pp.
  563--570, Feb. 2008.

\bibitem{boyer2013backscatter}
C.~Boyer and S.~Roy, ``{Backscatter communication and RFID: Coding, energy, and
  MIMO analysis},'' \emph{IEEE Trans. Commun.}, Mar. 2013.

\bibitem{biswas2021direct}
R.~Biswas, M.~U. Sheikh, H.~Yi{\u{g}}itler, J.~Lempi{\"a}inen, and
  R.~J{\"a}ntti, ``Direct path interference suppression requirements for
  bistatic backscatter communication system,'' in \emph{Proc. IEEE 93rd Veh.
  Technol. Conf. (VTC-Spring)}, Apr. 2021.

\bibitem{mollen2016uplink}
C.~Mollen, J.~Choi, E.~G. Larsson, and R.~W. Heath, ``{Uplink performance of
  wideband massive MIMO with one-bit ADCs},'' \emph{IEEE Trans. Wireless
  Commun.}, vol.~16, no.~1, pp. 87--100, Oct. 2016.

\bibitem{varshney2017lorea}
A.~Varshney, O.~Harms, C.~P{\'e}rez-Penichet, C.~Rohner, F.~Hermans, and
  T.~Voigt, ``{Lorea: A backscatter architecture that achieves a long
  communication range},'' in \emph{Proc. ACM Conf. Embedded Netw. Sensor
  Syst.}, Nov. 2017.

\bibitem{li2019capacity}
D.~Li, ``{Capacity of backscatter communication with frequency shift in Rician
  fading channels},'' \emph{IEEE Wireless Commun. Lett.}, vol.~8, no.~6, pp.
  1639--1643, Dec. 2019.

\bibitem{tao2021novel}
Q.~Tao, Y.~Li, C.~Zhong, S.~Shao, and Z.~Zhang, ``A novel interference
  cancellation scheme for bistatic backscatter communication systems,''
  \emph{IEEE Commun. Lett.}, vol.~25, no.~6, pp. 2014--2018, Jun. 2021.

\bibitem{zawawi2018multiuser}
Z.~B. Zawawi, Y.~Huang, and B.~Clerckx, ``{Multiuser wirelessly powered
  backscatter communications: Nonlinearity, waveform design, and SINR-energy
  tradeoff},'' \emph{IEEE Trans. Wireless Commun.}, vol.~18, no.~1, pp.
  241--253, Nov. 2018.

\bibitem{hua2020bistatic}
M.~Hua, L.~Yang, C.~Li, Z.~Zhu, and I.~Lee, ``{Bistatic backscatter
  communication: Shunt network design},'' \emph{IEEE Internet Things J.},
  vol.~8, no.~9, pp. 7691--7705, Nov. 2020.

\bibitem{guo2018exploiting}
H.~Guo, Q.~Zhang, S.~Xiao, and Y.-C. Liang, ``Exploiting multiple antennas for
  cognitive ambient backscatter communication,'' \emph{IEEE Internet Things
  J.}, vol.~6, no.~1, pp. 765--775, Jul. 2018.

\bibitem{tse2005fundamentals}
D.~Tse and P.~Viswanath, \emph{{Fundamentals of Wireless Communication}}.\hskip
  1em plus 0.5em minus 0.4em\relax Cambridge, U.K.: Cambridge Univ. Press,
  2005.

\bibitem{dobkin2012rf}
D.~Dobkin, \emph{{The RF in RFID}}, 2nd~ed.\hskip 1em plus 0.5em minus
  0.4em\relax New York, NY, USA: Newnes, 2013.

\end{thebibliography}

\end{document}